\documentclass[aps,prl,twocolumn,superscriptaddress,showpacs]{revtex4-2}

\def\F32{Fe$_3$Sn$_2$}
\def\cm{cm$^{-1}$}
\def\um{$\mu$m}

\usepackage{enumitem}
\usepackage{graphicx}
\usepackage{sidecap}
\sidecaptionvpos{figure}{t}
\usepackage{color}
\usepackage{epstopdf}
\usepackage{amssymb}
\usepackage{amsmath}
\usepackage{amsfonts}
\definecolor{darkblue}{rgb}{0,0.02,0.45}

\usepackage{color}
\usepackage[colorlinks,bookmarks=false,citecolor=darkblue,linkcolor=red,urlcolor=blue]{hyperref} 
\definecolor{darkred}{rgb}{0.7,0.0,0.0}

\definecolor{darkblue}{rgb}{0,0.02,0.45}

\definecolor{darkgreen}{rgb}{0.02,0.45,0.0}

\definecolor{violet}{rgb}{0.8,0.2,0.6}

\begin{document}

\title{High-pressure modulation of breathing kagome lattice: Cascade of Lifshitz transitions and evolution of the electronic structure}

\author{Marcos V. Gon\c{c}alves-Faria}
\thanks{These two authors contributed equally}
\affiliation{Helmholtz-Zentrum Dresden-Rossendorf, Ion Beam Physics and Materials Research, 01328 Dresden, Germany}
\affiliation{Institute of Applied Physics, TUD Dresden University of Technology, 01062 Dresden, Germany}

\author{Maxim Wenzel}
\thanks{These two authors contributed equally}
\affiliation{ 1. Physikalisches Institut, Universität Stuttgart, D-70569, Stuttgart, Germany}

\author{Yuk Tai Chan}
\affiliation{ 1. Physikalisches Institut, Universität Stuttgart, D-70569, Stuttgart, Germany}

\author{Olga Iakutkina}
\affiliation{ 1. Physikalisches Institut, Universität Stuttgart, D-70569, Stuttgart, Germany}

\author{Francesco Capitani}
\affiliation{ SMIS beamline, Synchrotron SOLEIL, L'Orme des Merisiers, Départementale 128, 91190 Saint-Aubin, France}

\author{Davide Comboni}
\affiliation{ESRF, BP 220, 38043 Grenoble Cedex 9, France}

\author{Michael Hanfland}
\affiliation{ESRF, BP 220, 38043 Grenoble Cedex 9, France}

\author{Qi Wang}
\affiliation{School of Physical Science and Technology, ShanghaiTech University, Shanghai 201210, China}
\affiliation{ShanghaiTech Laboratory for Topological Physics, ShanghaiTech University, Shanghai 201210, China}
\affiliation{ School of Physics and Beijing Key Laboratory of Opto-electronic Functional Materials \& Micro-nano Devices, Renmin University of China, Beijing 100872, China }

\author{Hechang Lei}
\affiliation{ School of Physics and Beijing Key Laboratory of Opto-electronic Functional Materials \& Micro-nano Devices, Renmin University of China, Beijing 100872, China }
\affiliation{Key Laboratory of Quantum State Construction and Manipulation (Ministry of Education), Renmin University of China, Beijing, 100872, China}

\author{Martin Dressel}
\affiliation{ 1. Physikalisches Institut, Universität Stuttgart, D-70569, Stuttgart, Germany}

\author{Alexander A. Tsirlin}
\affiliation{Felix Bloch Institute for Solid-State Physics, Leipzig University, 04103, Leipzig, Germany}

\author{Alexej Pashkin}
\affiliation{Helmholtz-Zentrum Dresden-Rossendorf, Ion Beam Physics and Materials Research, 01328 Dresden, Germany}

\author{Stephan Winnerl}
\affiliation{Helmholtz-Zentrum Dresden-Rossendorf, Ion Beam Physics and Materials Research, 01328 Dresden, Germany}

\author{Manfred Helm}
\affiliation{Helmholtz-Zentrum Dresden-Rossendorf, Ion Beam Physics and Materials Research, 01328 Dresden, Germany}
\affiliation{Institute of Applied Physics, TUD Dresden University of Technology, 01062 Dresden, Germany}

\author{Ece Uykur}
\email{e.uykur@hzdr.de}
\affiliation{Helmholtz-Zentrum Dresden-Rossendorf, Ion Beam Physics and Materials Research, 01328 Dresden, Germany}

\date{\today}

\begin{abstract}
The interplay between electronic correlations, density wave orders, and magnetism gives rise to several fascinating phenomena. In recent years, kagome metals have emerged as an excellent platform for investigating these unique properties, which stem from their itinerant carriers arranged in a kagome lattice. Here, we show that electronic structure of the prototypical kagome metal, \F32, can be tailored by manipulating the breathing distortion of its kagome lattice with external pressure. The breathing distortion is suppressed around 15~GPa and reversed at higher pressures. These changes lead to a series of Lifshitz transitions that we detect using broadband and transient optical spectroscopy. Remarkably, the strength of the electronic correlations and the tendency to carrier localization are enhanced as the kagome network becomes more regular, suggesting that breathing distortion can be a unique control parameter for the microscopic regime of the kagome metals and their electron dynamics.
\end{abstract}

\maketitle


\section*{INTRODUCTION}

Corner-sharing triangular geometry of the kagome lattice combined with itinerant carriers gives rise to various exotic phenomena, such as flat-band ferromagnetism~\cite{Mielke1992, Tasaki1992, Zhang2010}, Wigner crystallization~\cite{Wu2007, Wu2008}, fractional/anamolous quantum Hall effect~\cite{Neupert2011, Sheng2011, Regnault2011,Sankar2024}, Bose-Einstein condensation~\cite{Huber2010}, etc. The resulting interplay between magnetism, superconductivity, density-wave orders, and strong electronic correlations has been investigated by various experimental methods~\cite{Yin2022, Wang2023, Wilson2024}. Recent studies have revealed the presence of Dirac/Weyl states and flat bands~\cite{Morali2019, Liu2019, Kang2020, Liu2020, Kang2020a, Chen2023}, chiral spin structures~\cite{Ghimire2020}, and skyrmionic lattices~\cite{Hou2017, Hou2018} in magnetic kagome metals. In contrast, non-magnetic counterparts have shown intriguing density-wave orders and superconductivity~\cite{Ortiz2019, Ortiz2020, Jiang2022, Romer2022}. Several new compounds, such as FeGe, have also been discovered that bridge the gap between these two categories~\cite{Teng2022}. Owing to the wide variety of intriguing phenomena that can be realized in itinerant kagome systems, a natural question arises about how these properties are related to the underlying kagome network and whether they can be tuned with external stimuli. External pressure is especially valuable in this context, as it allows to tune the electronic structure and band filling without introducing additional chemical disorder.

\begin{figure*}
\centering
	\includegraphics[width=2\columnwidth]{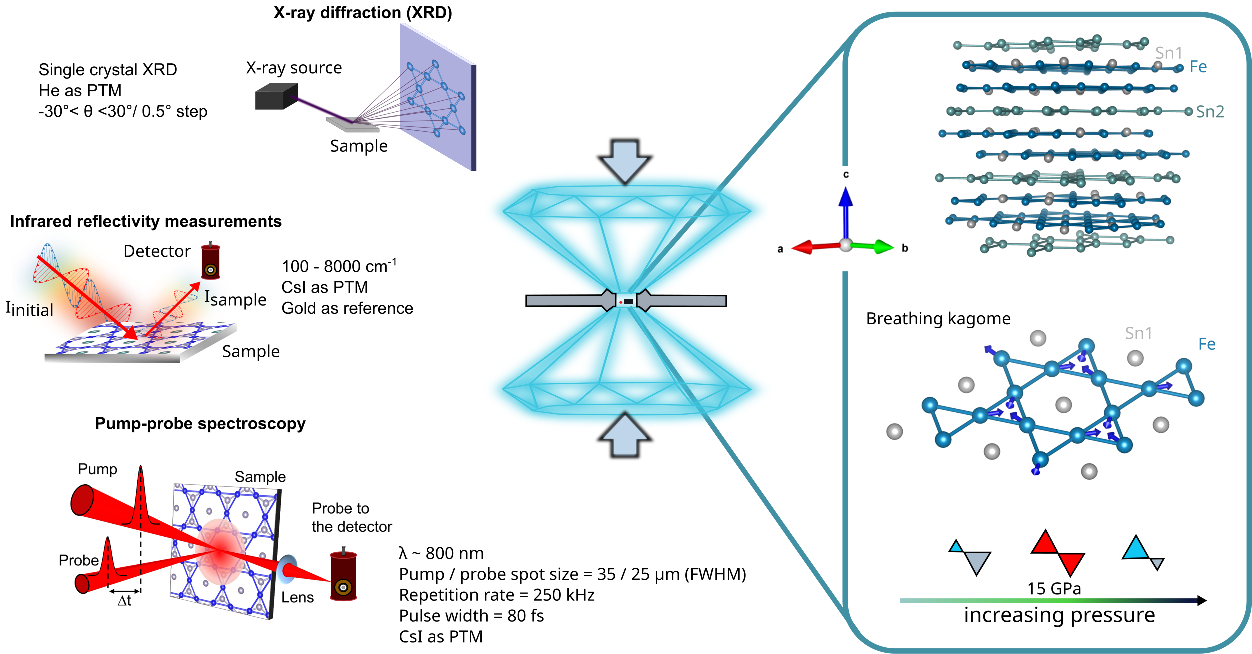}
	\caption{Summary of the experimental methods used in the present work: X-ray diffraction, broadband infrared spectroscopy, and optical pump-probe spectroscopy have been performed on single crystals prepared from the same crystal. Suitable pressure cells and pressure transmitting medium have been chosen for each method specifically. Bilayer kagome metal \F32\ is tuned with external pressure, leading to a gradual suppression and eventual reversal of the breathing distortion, accompanied by a cascade of Lifshitz transitions.}		
	\label{F1}
\end{figure*}

Compression of the kagome metals often affects the observer" atoms adjacent to the kagome network, whereas the most alluring kagome $d$-bands remain almost unchanged. For example, reentrant behavior of superconductivity in pressurized CsV$_3$Sb$_5$ is driven by the evolution of the Sb $p$-bands of the material without affecting its kagome $d$-bands~\cite{Tsirlin2021, Wenzel2024}. In this context, \F32\ is a rare example of the kagome lattice deformed by a breathing distortion~\cite{Tanaka2020} (Fig. 1) that couples to the external pressure and drives pressure-induced changes in the kagome bands, as we demonstrate in the following. At ambient pressure, \F32\ is a ferromagnet with the Curie temperature of 640 K~\cite{Kumar2019}. It features flat bands and massive Dirac fermions~\cite{Ye2018, Yin2018, Ye2019, Biswas2020}, a non-trivial Hall response~\cite{Du2022, Wang2024}, intriguing non-Fermi-liquid behavior~\cite{Ekahana2024}, and ultrafast spin-wave transport~\cite{Lee2023}. Little is known about its pressure evolution, though, beyond the mere fact that lattice parameters evolve smoothly upon compression without any abrupt structural phase transitions up to at least 20 GPa~\cite{Giefers2006}. On the other hand, within the R$\bar{3}$m space group, Fe, Sn1, and Sn2 sit at the atomic positions of (x,-x,z1), (0,0,z2), and (0,0,z3), respectively. The non-special atomic positions suggest that the motion of the Sn1 and Sn2 atoms is confined to the out-of-plane direction, whereas the Fe atoms can move within all three dimensions of the kagome network under external pressure creating a pathway towards the tunability of the breathing distortion. 

Here, we demonstrate the tunability of the kagome lattice of \F32\ via external pressure and the resulting electronic reconstruction traced by different optical studies (Fig.\ref{F1}). Structural evolution monitored by single-crystal x-ray diffraction (XRD) shows the suppression of the breathing distortion around 15~GPa and its reversal at higher pressures. These changes give rise to a series of Lifshitz transitions that fully reshape the Fermi surface of \F32\ and change its crucial parameters, such as the correlation strength, back-scattering of the unconventional carriers, and electronic relaxation. We show that the breathing mode of the kagome lattice is a powerful tool for tailoring the kagome $d$-bands and modifying electronic properties of these materials.

\section*{EXPERIMENTAL}

\noindent\textbf{Single Crystal X-ray Diffraction Measurements.} 
The structural study was performed on a single crystal of \F32. A diamond anvil cell (DAC) with 500 $\mu$m culet size was used in the measurements. The pressure transmitting medium was helium to ensure the high hydrostaticity in the pressure cell. Pressure values were calibrated using the ruby luminescence technique~\cite{Mao1986}. Single crystal X-ray diffraction (XRD) measurements were performed at room temperature in the ID15b beamline of ESRF. The sample-to-detector distance was calibrated using a Si standard and a vanadinite (Pb$_3$(VO$_4$)$_3$Cl) single crystal; the distance-to-detector was approximately 180~mm. Diffraction data were collected with the x-ray wavelength of 0.4099~\AA\ (convergent monochromatic beam with E $\approx$ 30 keV and $\sim$ 200 mA)) with 0.5$^\circ$ cell rotations about the $\theta$ axis. The data were integrated using Crysalis Pro~\cite{crysalispro}. About 500 reflections with 130 unique reflections were used to refine the isotropic atomic displacement parameters and the atomic coordinates of Fe, Sn1, and Sn2 in the R$\bar{3}$m symmetry. The refinements were performed in JANA2006~\cite{jana2006} against the data collected at pressures up to 28~GPa.\\

\noindent\textbf{Infrared Reflectivity Measurements.}
For the optical measurements, a sample with dimensions of 150$\mu$m $\times$ 150$\mu$m $\times$ 50$\mu$m was prepared from the as-grown single crystal. An as-grown surface is used for the experiments. High-pressure reflectivity measurements were performed at room temperature at the SMIS beamline of the SOLEIL synchrotron, France, on a homemade horizontal microscope with custom Schwarzschild objectives (NA = 0.5). A diamond anvil cell (DAC) with type-IIa diamond anvils and a culet of 600\um\ diameter was utilized. CsI powder is used as the pressure-transmitting medium, ensuring a good diamond-sample interface~\cite{Celeste2019}. The sample and Ruby spheres used as pressure gauges were placed inside a stainless steel gasket with a 200\um\ diameter hole. The pressure was determined by monitoring the calibrated shift of the ruby R1
fluorescence line as described in Ref.~\cite{Mao1986}. The reflectivity spectra at the sample-diamond interface were recorded in a broad spectral range of 150 - 10000~\cm\ by a Thermo-Fisher iS50 interferometer with KBr and solid substrate beamsplitters, using an MCT detector and a liquid helium-cooled bolometer. The reflectivity of a gold foil loaded into the DAC at ambient pressure served as a reference. Other optical quantities were obtained using standard Kramers-Kronig (KK) analysis considering the sample-diamond interface as explained in SM. \\

\noindent\textbf{Ultrafast Optical Pump-Probe Measurements and Raman Measurements.} 
Pressure-dependent transient reflectivity and Raman spectroscopy were measured using the same diamond anvil cell (DAC). In both cases, small pieces of \F32\ (around 150-200 $\mu$m diameter and 30-40 $\mu$m  thickness) single crystals were placed inside an Almax Plate DAC (49 mm diamond anvil diameter and 10.5 mm working distance to the sample). The pressure inside the DAC was monitored using the standard ruby fluorescence~\cite{Mao1986}. CsI powder was used as the pressure-transmitting medium for the time-resolved experiments to ensure direct contact between the sample surface and the diamond. For Raman measurements, silicon oil was the pressure-transmitting medium.\\
• Optical pump-probe: High-pressure optical pump-probe experiments up to 12~GPa were performed using the reflection geometry. For the pump and probe, we used laser pulses centered at $\lambda$ = 800 nm, with a pulse duration of around 80 fs. The setup is based on a Ti: sapphire laser amplifier operating with a 250 kHz repetition rate. Pump and probe spots were focused on the sample surface inside the DAC and were around 35 µm and 25 µm (FWHM) diameter, respectively. Pump-probe traces were measured up to 130 ps, with a finer time steps during the first 8 ps to resolve the coherent phonon.\\
• Raman spectroscopy:
Raman experiments up to 18~GPa were performed with a HORIBA commercial system. A 532~nm laser was used for the incoming beam to excite the compound. The laser is focused on the sample surface with a 50x magnification objective, and the spot size was around 5~$\mu$m diameter. The contributions from the diamond (starting around 160~\cm) and the elastic scattering at lower wavenumbers were subtracted from the original data to focus on the A$_{1g}$ phonon.\\

\noindent\textbf{Calculation Details.}
Full-relativistic DFT calculations were performed in the FPLO (magnetic moments)~\cite{fplo} and Wien2K (Fermi surface and optical response)~\cite{Blaha2020,wien2k} and VASP (equation of state, phonons)~\cite{Kresse1999} codes using Perdew-Burke-Ernzerhof (PBE) exchange-correlation potential~\cite{pbe96} and the $24\times 24\times 24$ $k$-mesh that ensures convergence of the Fermi energy. Lattice parameters and atomic positions were fixed to their experimental values determined from our XRD experiments. Spin-orbit coupling was included in the calculations, and the ferromagnetic order of the sample has been taken into account by assuming the spins aligned along the $c$-axis are in line with the experimental magnetic structure at ambient pressure and room temperature. The obtained total magnetic moment per Fe at ambient pressure is $\sim$2.05~$\mu_B$, consistent with the experimental results~\cite{Wang2016}. Optical conductivity was calculated using the optic module~\cite{AmbroschDraxl2006} on the denser $k$-mesh with up to 50$\times$50$\times$50 points. Frequencies of $\Gamma$-point phonons were obtained from the built-in procedure with frozen atomic displacements of 0.015 $\AA$. \\

\section*{RESULTS AND DISCUSSION}
\subsection{Structural evolution of the breathing mode}

\begin{figure*}
\centering
	\includegraphics[width=2\columnwidth]{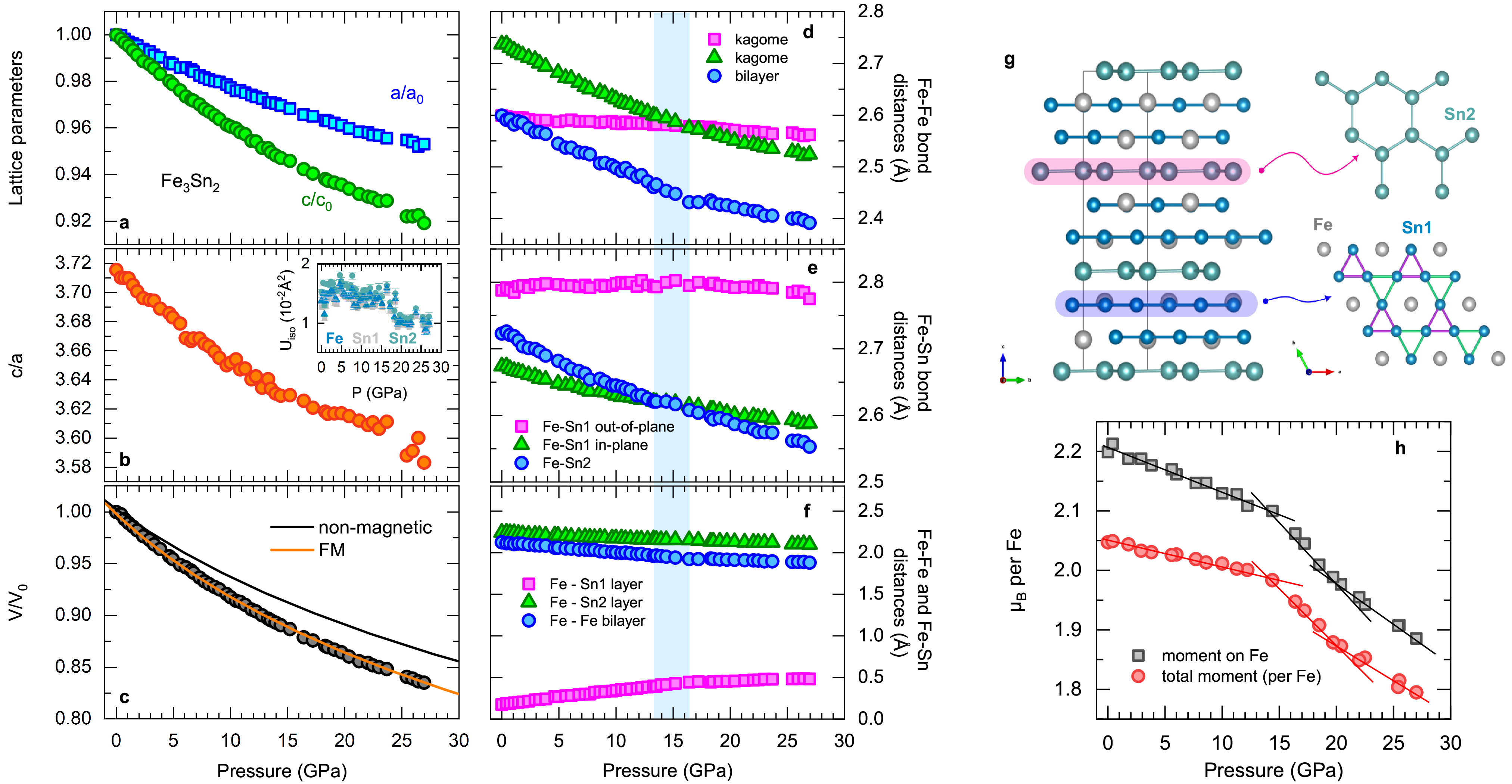}
	\caption{ Pressure-dependent structural parameters of \F32. (a-c) Pressure-dependent lattice parameters and the volume change. Inset in \textbf{b} shows atomic displacement parameters of Fe, Sn1, and Sn2. Overall, all the atoms show similar displacement with a slight decrease and a sudden drop at the high-pressure range. That is expected as the system becomes denser. The black and orange lines correspond to the equations of state obtained from the DFT calculations for the nonmagnetic and ferromagnetic cases, respectively. Only the ferromagnetic calculations reproduce the experimental behavior. (d) Fe-Fe bond distance between two layers of the bilayer is gradually reduced. By contrast, the Fe-Fe distances within the same layer of the kagome network (color code is given in \textbf{g}) cross around 15~GPa, indicating the suppression and eventual reversal of the breathing distortion. (e) Fe-Sn bond distances. The Fe-Sn1 (out of plane) does not change significantly, suggesting that the overall bilayer structure is robust. (f) Fe-Fe and Fe-Sn distances (vertical). Fe-Sn1 shows an unexpected increase as the Sn1 layers are slightly pushed away from the kagome network. As discussed in the main text, this also has certain effects on the phonon dynamics in \F32. (g) Crystal structure of \F32. The breathing mode is also shown with a color code. (h) Calculated magnetic moment (total moment per Fe and moment on Fe). The total moment of 2.05 $\mu_B$ per Fe at ambient pressure is consistent with the literature~\cite{Wang2016}. The kink at higher pressures marks the Lifshitz transitions, as discussed in the text.}		
	\label{XRD}
\end{figure*}

In order to identify the structural changes, in particular the pressure evolution of the breathing mode, we performed a single crystal XRD study up to $\sim$ 28~GPa. No structural phase transition is observed in agreement with the previous powder XRD studies~\cite{Giefers2006}. As demonstrated in Fig.~\ref{XRD} (a-c), both $a$ and $c$ lattice parameters evolve gradually, with only a slightly more pronounced change along the $c$-axis, as one might expect from a system with a layered structure like \F32. Both magnetic and nonmagnetic $ab$ $initio$ calculations were used to estimate the compressibility of \F32. Experimental pressure dependence of the unit-cell volume is in an excellent agreement with the calculation for the ferromagnetic equation of state (see SM for the fit parameters), thus indirectly suggesting that ferromagnetism is preserved up to at least 28 GPa.
 
Two main features of the kagome structure in \F32\ are the formation of bilayers and the breathing distortion of each kagome layer therein. We track experimentally the evolution of the respective Fe-Fe distances (Fig.~\ref{XRD}(d-f)). The Fe-Fe bond distances decrease continuously following the lattice parameters within the bilayer structure, signalling that the coupling between these two layers is effectively increasing. The Fe-Fe distances within one kagome layer, on the other hand, reveal an interesting development. The Fe-Fe bonds on the consecutive corner-sharing triangles of the kagome structure do not shrink equally; rather, the long bond decreases faster. This leads to the gradual suppression of the breathing kagome distortion until the kagome layer becomes regular at $\sim$15~GPa. The breathing distortion is then reversed upon further compression. These results demonstrate that the external pressure can be efficiently used to tune the kagome lattice directly. As discussed below, approaching the perfect kagome structure has significant consequences for the electronic structure of \F32.

\subsection{Pressure-induced Lifshitz transitions}

\begin{figure*}
\centering
	\includegraphics[width=2\columnwidth]{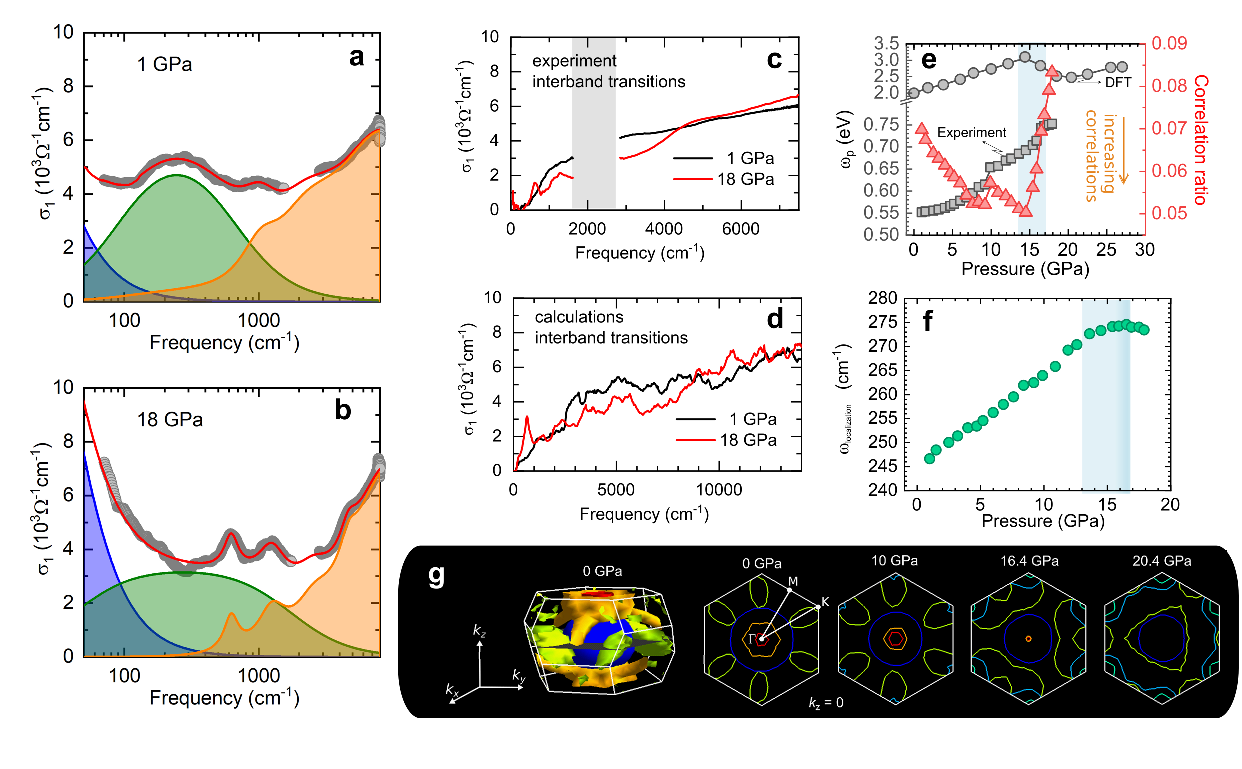}
	\caption{(a-b) Decomposition of the optical conductivity at the lowest and highest pressures measured at room temperature. Drude (blue), unconventional carriers (green), and the interband transitions (orange) are identified. (c-d) Comparison of the experimental and calculated interband transitions at 1- and 18~GPa. There are no significant changes with pressure, but subtle changes can be reproduced very well. The energy scale of the calculated optical conductivity is rescaled by $\sim$/2 to account for band energy renormalization in a correlated system. (e) Experimental and calculated plasma frequencies. The correlation ratio is also given on the same plot. The tendency of increasing the correlations is reversed above $\sim$15~GPa. (f) Pressure-dependent resonance of the localization peak gradually increases and eventually saturates above $\sim$15~GPa. (g) Fermi surfaces at different pressures. Chosen pressures represent the series of Lifshitz transitions.}		
	\label{IR}
\end{figure*}

Next we utilized optical spectroscopy to study the electronic changes accompanying the structural evolution in \F32. Previous optical studies at ambient pressure have revealed several interband transitions as well as conventional and unconventional intraband transitions in \F32~\cite{Biswas2020}. All of these features can be traced under pressure. In Fig.~\ref{IR}(a,b), an exemplary decomposition of the low- and high-pressure optical conductivity is given, showing a conventional Drude contribution (blue), an unconventional localization peak (green), and several interband transitions (orange). Here we applied the same decomposition method used for the other kagome metals as explained in Ref.~\cite{Wenzel2024}. 

More pronounced low-energy interband transitions appear in the spectra with increasing pressure, while simultaneously, the MIR spectral weight is slightly suppressed. Even though the changes in the interband transitions are not very significant, they can be used to benchmark our DFT calculations. In Fig.~\ref{IR}(c,d), a comparison of the optical spectra is given for 1~ and 18~ GPa. The intraband transitions (namely Drude+localization) have been removed from the response to facilitate a direct comparison between the experimental interband transitions of the optical response and the DFT calculations. The suppression of the MIR spectral weight, the low energy absorptions, and the crossover energies between low- and high-pressure spectra are well reproduced by the DFT calculations upon a suitable energy renormalization. This mismatch in the energy scales is not very surprising since \F32\ is a strongly correlated electron system~\cite{Yin2018}. However, these results indicate that the DFT calculations can be used as a guide to the changes in the electronic structure. 

Intraband part of the optical conductivity gives further insights into the pressure dependence of the electronic structure. While taking into account the Drude and the localization peak, we determined the plasma frequency ($\omega_p^2 \approx$ carrier density) in \F32 (Fig~\ref{IR}(e)). An overall increase in the plasma frequency is observed, which is in line with the increase in the Fermi surface size. A closer look reveals non-monotonous changes at $\sim$10~ and 16~GPa. Sudden jumps are indications of reformation in the Fermi surface topology, the Lifshitz transitions~\cite{Pietro2018}. Indeed the Fermi surfaces plotted as a function of pressure (Fig.~\ref{IR}(g)) reveal the appearance of new sheets at the $K$-point of the Brillouin zone at these critical pressures, demonstrating the close link to the observed changes in the plasma frequency.  

Another indication of the Lifshitz transition is the kink in the pressure dependence of the magnetic moment around 15~GPa. The magnetic moment corresponds to the difference in the number of states in the spin-up and spin-down channels at the Fermi level, so it also reflects an abrupt change in the Fermi surface when the breathing distortion is suppressed. (Fig.~\ref{XRD}(h)). Further modifications of the Fermi surface are observed at higher pressures. Two Fermi surfaces at the $\Gamma$-point disappear when the breathing distortion is reversed. 

The direct comparison of the experimental plasma frequencies to those calculated from the DFT calculations can give an insight into the correlation strength in materials~\cite{Shao2020} (See SM for details). At ambient pressure, \F32\ stands out as one of the most correlated compounds among the different classes of kagome metals (For comparison see~\cite{Shao2020, Wenzel2022}). Our pressure-dependent study shows that \F32\ reaches its most correlated state with the suppression of the breathing distortion. Above this crossover, the correlation strength decreases; however, up to at least $\sim$ 18 GPa, it remains highly correlated. The non-monotonic change in correlation strength demonstrates that multiple mechanisms are at play regarding the electronic correlations in \F32, and it strongly indicates the switch of the dominant contribution under pressure. The significant alteration of the Fermi surface at around 15~GPa notably affects these correlations, as well.

\subsection{Localized carriers}

The localization peak reflects the unconventional carriers in kagome metals and is commonly observed in different members of this materials class~\cite{Wenzel2022, Wenzel2024}. While sometimes this response can be identified as a second Drude contribution reflecting charge carriers with different scattering rates~\cite{Schilberth2022, Mosesso2024}, in most cases including \F32, a displaced Drude peak due to strong localization effects is observed. These localized carriers arise from the interactions of charge carriers with low-energy excitations, such as phonons and electric or magnetic fluctuations, leading to a back-scattering of the electrons~\cite{Fratini2014, Fratini2021}. Although different classes of kagome metals exhibit these carriers, the temperature and pressure-dependent features depend on the details of the studied materials. However, the general expectation is that increasing pressure should delocalize these carriers, and the localization peak should shift to the lower energies, evolving into a Drude-like contribution, and/or disappear. Indeed, such a behavior is observed upon the initial compression of the nonmagnetic kagome metal CsV$_3$Sb$_5$~\cite{Wenzel2023}. 

In \F32, on the other hand, the localization peak shifts to higher energies (Fig.~\ref{IR}(f)) contrary to the expectations right before it saturates above the 15~GPa crossover. This behavior suggests that the carrier localization in the kagome metals may be related to the kagome geometry itself, because it increases as the kagome network becomes more regular upon compression.

\subsection{Non-equilibrium electron dynamics}

\begin{figure*}
\centering
	\includegraphics[width=2\columnwidth]{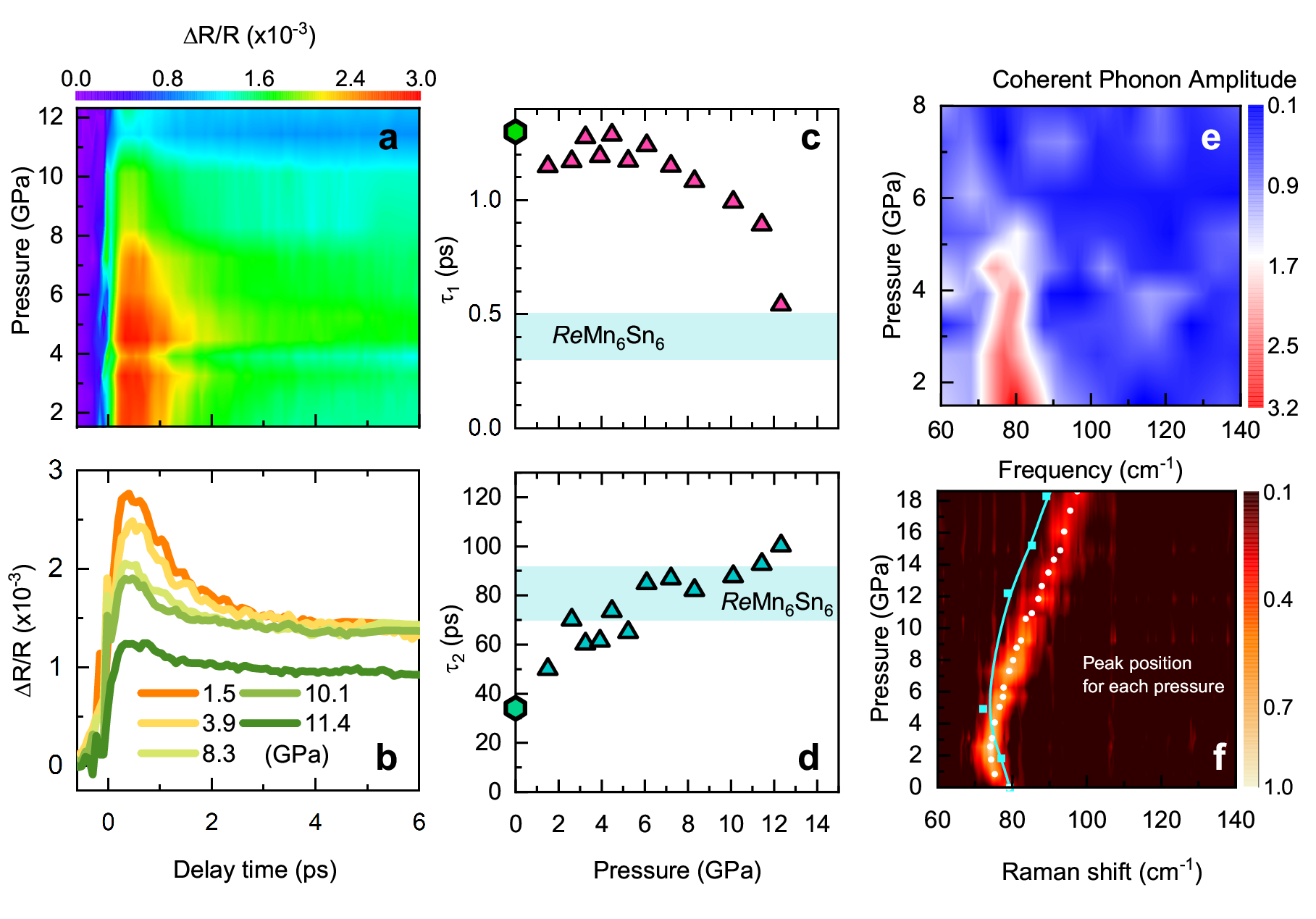}
	\caption{ (a-b)Pressure-dependent transient reflectivity. (c-d) Relaxation parameters are obtained via the exponential fit of the transient reflectivity (see SM for the details), where $\tau_1$ and $\tau_2$ represent the conventional and unconventional carriers, respectively. Ambient pressure data is from Ref.~\cite{GoncalvesFaria2024} (e) Pressure-dependent coherent phonon amplitude. A 2.4~THz phonon is identified in the transient reflectivity and disappears above $\sim$4-5~GPa. (f) Pressure-dependent Raman shift of the 2.4~THz $A_{1g}$ phonon. A phonon softening up to $\sim$4~GPa is observed, followed by the gradual hardening of the mode. White points are the resonance frequency of the mode obtained with a Lorentzian fit (see SM for details). Blue points are the calculated phonon frequency via DFT. A softening also observed in DFT suggests the lattice tendency for the softening up to around 5 GPa.  }		
	\label{PP}
\end{figure*}

It is further instructive to address the behavior of charge carriers in the kagome metal beyond equilibrium; therefore, we expanded our investigation to ultrafast optical pump-probe measurements under pressure. At ambient pressure, one observes coherent phonon oscillations and two distinct relaxation dynamics in this material~\cite{GoncalvesFaria2024}. A short relaxation time of around 1~ps was attributed to the hot electron cooling ($\tau_1$). A second relaxation time on the order of 10s of ps ($\tau_2$) was also observed, which is clearly distinct from the heat dissipation due to the lattice cooling, shown to be extremely slow (on the order of ns). The short relaxation time is expected from a metallic system such as \F32. The presence of the second relaxation time in such long time scale is clearly unexpected and can be associated with the presence of unconventional, localized carriers. Two relaxation processes were also observed in other kagome metals~\cite{Tuniz2023,Wang2021,Liu2023}.

Our high-pressure optical pump-probe study, as summarized in Fig.~\ref{PP} (see SM for the details), tracks changes in electron dynamics upon reducing the breathing distortion of the kagome lattice. With increasing pressure, $\tau_1$ evolves non-monotonically. Up to around 5~GPa, the relaxation time is almost constant, whereas, above 5~GPa, a strong suppression is observed along with the amplitude. This sudden change in $\tau_1$ likely indicates the change of electron-phonon coupling as the fast relaxation process is governed by the electron-phonon coupling strength. $\tau_2$, on the other hand, increases significantly with increasing pressure without any pronounced anomalies. With increasing pressure, both relaxation processes approach the values observed in the closely related $Re$Mn$_6$Sn$_6$ systems~\cite{Faria2023}. These compounds possess well separated, non-distorted kagome layers. The fact that the relaxation dynamics in \F32\ approach those of $Re$Mn$_6$Sn$_6$ within the pressure range where the breathing mode is suppressed is particularly interesting and indicates the universal behavior of the electron dynamics in the kagome metals signaling to the intricate interplay between the kagome network and the relaxation dynamics.

External pressure has a significant impact on coherent phonon oscillations, as well. Robust phonon oscillations at ambient pressure are ascribed to the 2.4 THz A$_{1g}$ mode with the out-of-plane displacements of the Sn1 atoms that center the hexagons of the kagome network. It is presently unclear why this mode is strongly coupled to the electronic background, even though it does not change the positions of Fe atoms. However, the Fe atoms may be affected indirectly, because Sn1 stabilizes the kagome network. 

With increasing pressure, the coherent phonon oscillations weaken and gradually disappear above around 5~GPa (Fig.~\ref{PP}(e)). Interestingly, this pressure also corresponds to the pronounced suppression of the fast relaxation process, further evidencing the unconventional interplay between electrons and phonons in \F32. While the energy resolution of our pump-probe experiments is not good enough to resolve the exact frequency dependence of this mode, a Raman study (details are given in SM) clearly shows the non-monotonic behavior (Fig.~\ref{PP}(f)). The 2.4~THz A$_{1g}$ Sn1-mode softens with increasing pressure up to around 4-5~GPa, followed by the expected phonon hardening at higher pressures. This behavior is well reproduced by our DFT calculations (Fig.~\ref{PP}(f), blue dots). It can be ascribed to the two competing effects. On the one hand, pressure shifts Sn1 out of the Fe kagome plane and makes the out-of-plane displacements easier. On the other hand, compression along '$c$' increases the frequencies of the out-of-plane modes. Therefore, the behavior of the A$_{1g}$ mode itself is not anomalous, but its coupling to the electronic background is apparently reduced as the Sn1 atoms move out of the kagome plane under pressure. 

\section*{CONCLUSIONS}
Compression of \F32\ has a direct impact on its electronic structure by modifying the breathing distortion of the kagome lattice. Increasing pressure leads to the suppression of the breathing distortion around 15~GPa and to its eventual reversal at higher pressures. This fundamental change in the structure leads to a cascade of Lifshitz transitions evidenced by several experimental signatures: i) anomaly in pressure dependence of the magnetic moment; ii) non-monotonic changes in the plasma frequency and correlation strength; iii) enhancement of the localized carriers; iv) non-monotonic changes in the relaxation dynamics and transient reflectivity. Using DFT calculations benchmarked by a direct comparison to the experimental optical conductivity, we identify two Lifshitz transitions with the appearance of a new Fermi surface sheet near $K$ around 10~GPa and the disappearance of the sheets near $\Gamma$, along with the merge of the pockets near the zone boundary, around 15~GPa. The suppression of the breathing distortion fully re-shapes the Fermi surface and strongly modifies electron dynamics in the kagome metal. Another reconstruction of the Fermi surface is expected at even higher pressures where the breathing distortion is reversed. This regime may be particular unusual, because it is characterized by the same correlation strength as at ambient pressure, but with the substantially increased carrier concentration. The breathing mode of the kagome network is a powerful tool for modifying the electronic structure and opens exciting prospects for manipulating kagome metals using pressure and strain.


\section*{Acknowledgements}

We are grateful to Gabriele Untereiner for preparing the single crystals for the high-pressure measurements. We also acknowledge the technical support by Uta Lucchesi during our Raman measurements. We thank SOLEIL synchrotron, France, for providing the beamtime (proposal No. 20210399) for high-pressure infrared measurements and ESRF syncrothron, Grenoble, for providing the beamtime (proposal No HC-4451) for high-pressure XRD measurements. H.L. is supported by the National Key R\& D Program of China (Grant Nos. 2022YFA1403800 and 2023YFA1406500), the National Natural Science Foundation of China (Grant No. 12274459). M.W. is supported by IQST Stuttgart/Ulm via a project funded by Carl Zeiss foundation. The work has been supported by the Deutsche Forschungsgemeinschaft (DFG) via UY63/2-1. Computations for this work were done (in part) using resources of the Leipzig University Computing Center.

\bibliography{F32HP.bib}

\newpage
\makeatletter
\renewcommand\@bibitem[1]{\item\if@filesw \immediate\write\@auxout
    {\string\bibcite{#1}{S\the\value{\@listctr}}}\fi\ignorespaces}
\def\@biblabel#1{[S#1]}
\makeatother
\setcounter{figure}{0}
\renewcommand{\thefigure}{S\arabic{figure}}
\newcommand{\av}{\mathbf a}
\newcommand{\bv}{\mathbf b}
\newcommand{\cv}{\mathbf c}

\clearpage
\newpage
\onecolumngrid
\begin{center}
\large{\textbf{\textit{Supplemental Material} \\ High-pressure modulation of breathing kagome lattice: Cascade of Lifshitz transitions and evolution of the electronic structure}}\\
\end{center}
\begin{center}
\large{Marcos V. Gon\c{c}alves-Faria$^{1, 2, \dagger}$, Maxim Wenzel$^{3, \dagger}$, Yuk Tai Chan$^{3}$, Olga Iakutkina$^{3}$, Francesco Capitani$^{4}$, Davide Comboni$^{5}$, Michael Hanfland$^{5}$, Qi Wang$^{6,7,8}$, Hechang Lei$^{8,9}$, Martin Dressel$^{3}$, Alexander A. Tsirlin$^{10}$, Alexej Pashkin$^{1}$, Stephan Winnerl$^{1}$, Manfred Helm$^{1, 2}$, Ece Uykur$^{1,*}$}\\
\end{center}

\subsection{Optical variables under pressure}

In our high-pressure infrared spectroscopy measurements, the reflectivity spectrum of \F32, R$_{sd}(\omega)$ is measured at the sample-diamond interface in the diamond anvil cell
(DAC) which is expressed as 

\begin{equation}
 R_{sd}(\omega) = R_{gd}(\omega)\frac{I_{sd}(\omega)}{I_{gd}(\omega)}
\end{equation}

Here, I$_{sd}(\omega)$ is the reflected light intensity at the sample-diamond interface, and I$_{gd}(\omega)$ is at the gold-diamond interface. At each pressure, the reflectivity of the sample has been corrected for the gold foil reflection with the R$_{gd}(\omega)$. Thus, the above equation gives us the absolute reflectivity R$_{sd}(\omega)$ obtained at the sample-diamond interface.

During the Kramers-Kronig (KK) analysis of the reflectivity at the sample-diamond interface, the standard relation between the phase and the reflectivity needs to be corrected, which gives an additional term (second term in the right-hand side of Eq. 2). Namely, the presence of a medium with the refractive index larger than 1 (in the case of diamond it is 2.38) brings an extra phase shift.

\begin{equation}
\Theta(\omega_0) = -\frac{\omega_0}{\pi}P\int_0^{+\infty}\frac{lnR_{sd}(\omega)}{\omega^2-\omega_0^2}d\omega+\left[\pi-2\text{arctan}\frac{\omega_\beta}{\omega_0}\right]
\end{equation}

$P$ is the principal value of the complex response function. $\omega_\beta$ is the position of the reflectivity pole on the imaginary axis of the complex plane. For the case of measurements performed at the sample-vacuum interface, $\omega_\beta$ goes to infinity, eliminating the second term and leading to the standard KK analysis.

$\omega_\beta$ is $a$ $priori$ an unknown parameter; however, it can be estimated by comparing the lowest-pressure optical conductivity (in our case, 1~GPa spectrum is used) with the ambient-pressure one, which can be obtained through the regular KK analysis. We found the best fitting for the parameter $\omega_\beta$ $\sim$ 25.000~\cm.

\begin{figure*}
\centering
	\includegraphics[width=1\columnwidth]{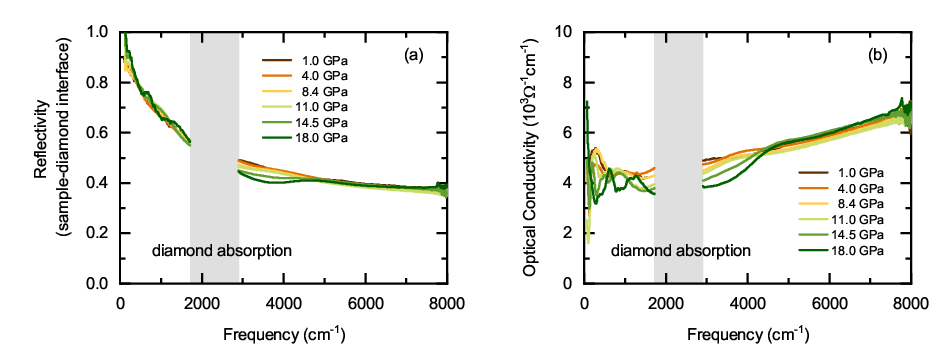}
	\caption{ (a) Pressure-dependent reflectivity and (b) optical conductivity. Shaded areas are diamond absorptions. }		
	\label{Optical}
\end{figure*}

In Fig.~\ref{Optical}, pressure-dependent reflectivity and the obtained optical conductivity are given.

\subsection{Electronic correlations}

The strength of the electronic correlations can be assessed by the comparison of the plasma frequencies obtained via experiment and DFT with the following equation:

\begin{equation}
\text{correlation ratio} = \frac{\omega_p^2 (\text{experiment})}{\omega_p^2 (\text{DFT})}
\end{equation}

The experimental plasma frequency can be calculated using the spectral weight (SW) of the intraband transitions, which includes contributions from both Drude and localization. In this study, we modeled our optical conductivity considering Drude, localization, and interband transitions. After subtracting the interband transitions, we determined the total spectral weight of the intraband transitions using the following equation:

\begin{equation}
\omega_p^2 = SW = \frac{120}{\pi} \int_0^{\omega_c} \sigma_1(\omega)d(\omega)
\end{equation}  

Here, $\omega_c$ is chosen as the upper limit of the measurement window, which is enough to cover the contributions from Drude and localization. 

\newpage
\subsection{ Details of pump-probe measurements and analysis}

Figure \ref{PP}(a) shows the transient reflectivity data at 1.5 GPa until 60 ps. The equation used to fit the traces is also shown in this panel. The same function was used for all pressure points, which has also been used before for \F32 outside the pressure cell. Panels (b) and (c) from Figure~ \ref{PP} show the pressure evolution of the fitting parameters C$_1$ and C$_2$, respectively. C$_1$ corresponds to the amplitude at time zero of the faster relaxation process (green line in panel (a)). It is attributed to the relaxation of hot electrons through electron-phonon interactions, which is the expected metallic behavior for this experiment. Interestingly, C$_1$ reduces drastically with increasing pressure, indicating the excitation of fewer charge carriers at higher pressures. C$_2$, on the other hand, shows a more constant behavior as a function of pressure. The origin of this longer relaxation process (blue line in panel (a)) is still not fully understood, despite being present in many of the kagome metals that have been measured with similar experiments~\hyperlink{cite}{\color{blue}[S1-S3]}. 

\begin{figure*}[h!]
\centering
	\includegraphics[width=1\columnwidth]{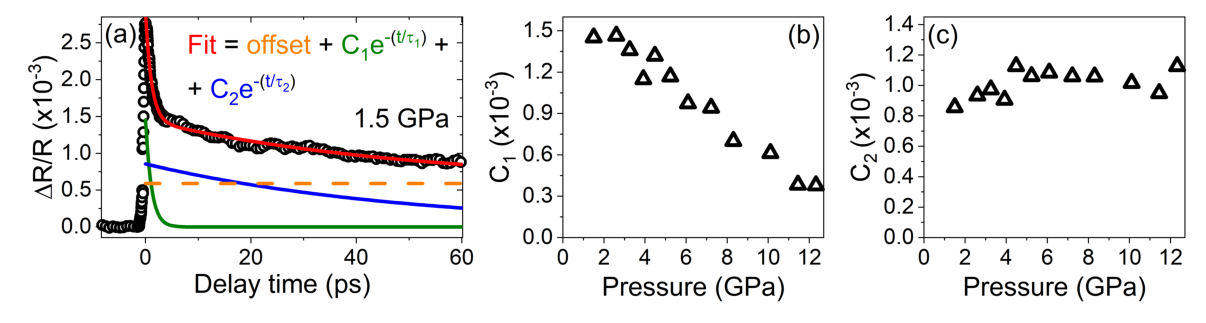}
	\caption{ (a) Fit procedure of the transient reflectivity spectra. The relaxation times are given in the main text Figure 4(c,d). (b,c) The amplitude of the different relaxation times as described in the equation given in (a). }		
	\label{PP}
\end{figure*}

\begin{figure*}
\centering
	\includegraphics[width=0.8\columnwidth]{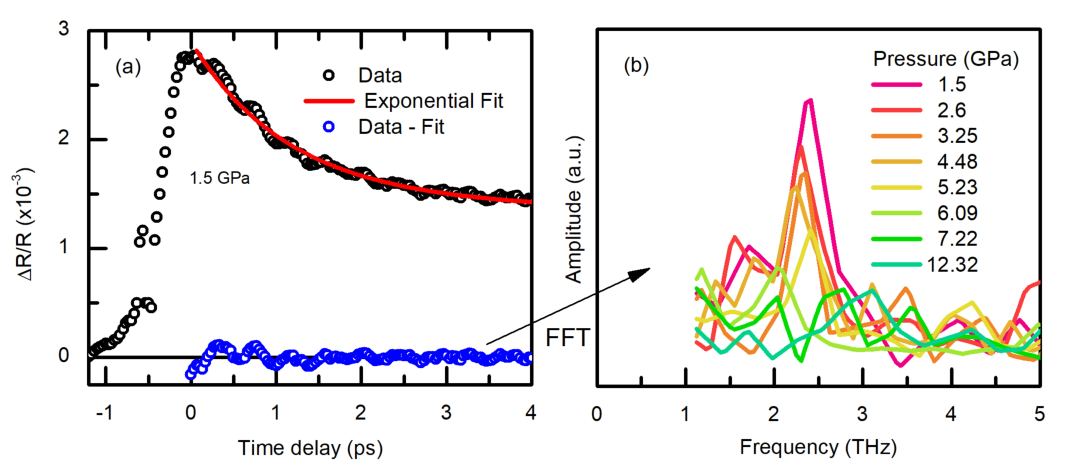}
	\caption{ (a) Obtaining the coherent phonon signal. (b) The amplitude of the Fourier transform of the oscillatory signal signifying the phonon mode that disappears above $\sim$5~GPa. }		
	\label{CP}
\end{figure*}

Coherent phonon is obtained following the procedure described in Figure~\ref{CP}. The double exponential fit of the transient reflectivity is subtracted from the pump-probe trace. The frequency of the phonon mode can be obtained via the Fourier transform of the remaining oscillatory signal. As seen from Figure~\ref{CP}(b), the amplitude of the phonon mode is gradually suppressed and disappears above $\sim$5~GPa within the noise level of our experiments.

\newpage
\subsection{ High-pressure Raman measurements}

\begin{figure*}[h!]
\centering
	\includegraphics[width=1\columnwidth]{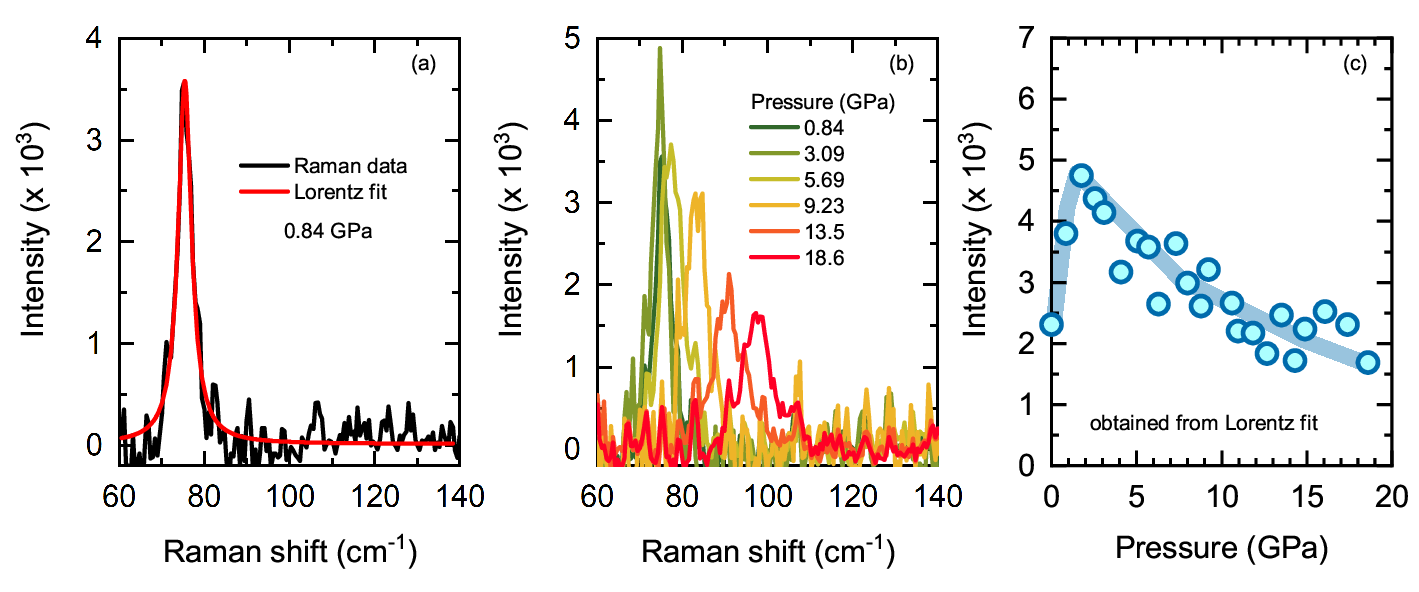}
	\caption{(a) Lorentzian fit of the Raman data. The same fitting was used for all the pressure points. (b) Pressure evolution of the A$_{1g}$ mode. (c) The pressure-dependent intensity of the phonon mode obtained by the Lorentzian fit. }		
	\label{Ramanfit}
\end{figure*}

In order to follow the phonon behavior at higher pressures and compare it with the optical pump-probe results presented in the main text, we performed high-pressure Raman spectroscopy. The white points in Figure 4(f) in the main text represent the peak position obtained after fitting the Raman data, as shown in Figure~\ref{Ramanfit}(a). The same fitting procedure was used for all the pressure points.

\newpage
\subsection{Structure refinement parameters and equation of state}

Fits of the energy-vs-volume curves with the Murnaghan equation of state~\hyperlink{cite}{\color{blue}[S4, S5]}  return the equilibrium volume $V_0=489.379(2)$\,\r A$^3$/f.u., bulk modulus $B_0=92.78(9)$\,GPa, and pressure derivative of the bulk modulus $B_0'=4.75(8)$.

\begin{table}[h!]
\centering
\caption{Details of data collection and refined structural parameters for \F32\ at 1~GPa.}
\begin{tabular}{c}\hline
 $a=b=5.3194(3)$\,\r A,\quad $c=19.734(13)$\,\r A \\
 $V=483.6(3)$\,\r A$^3$ \\
 $R\bar{3}m$ \\
 $\lambda=0.40990$\,\r A \\
 $2\theta_{\min}=2.62^{\circ}$, \quad $2\theta_{\max}=21.43^{\circ}$ \\
 $-9\leq h\leq 9$,\quad $-8\leq k\leq 8$,\quad $-11\leq l\leq 8$ \\
 Number of total/unique reflections = $508/132$\\
 $R_{I>3\sigma(I)}=0.028$, \quad $R_{I>3\sigma(I)}=0.0353$  \smallskip\\\hline
\end{tabular} 
\\\medskip
\begin{tabular}{c@{\hspace{1cm}}c@{\hspace{0.4cm}}c@{\hspace{0.4cm}}c@{\hspace{0.5cm}}c}
 Atom & $x/a$ & $y/b$ & $z/c$ & $U_{\rm iso}$ (\r A$^2$) \\
 Fe & 0.49581(9)  & 0.50418(3)   &  0.1131(1)   & 0.0114(3) \\
 Sn1 & 0 & 0 & 0.1034(1)  & 0.0127(3) \\
 Sn2 & 0  & 0   & 0.3316(1) &0.0151(3) \smallskip\\\hline
\end{tabular}
\label{Fe32_1GPa}
\end{table}

\begin{table}[h!]
\centering
\caption{Details of data collection and refined structural parameters for \F32\ at 26~GPa.}
\begin{tabular}{c}\hline
 $a=b=5.0858(10)$\,\r A,\quad $c=18.22(4)$\,\r A \\
 $V=408.2(9)$\,\r A$^3$ \\
 $R\bar{3}m$ \\
 $\lambda=0.40990$\,\r A \\
 $2\theta_{\min}=2.74^{\circ}$, \quad $2\theta_{\max}=20.31^{\circ}$ \\
 $-8\leq h\leq 8$,\quad $-8\leq k\leq 8$,\quad $-8\leq l\leq 9$ \\
  Number of total/unique reflections = $421/102$\\
 $R_{I>3\sigma(I)}=0.049$, \quad $R_{I>3\sigma(I)}=0.056$  \smallskip\\\hline
\end{tabular} 
\\\medskip
\begin{tabular}{c@{\hspace{1cm}}c@{\hspace{0.4cm}}c@{\hspace{0.4cm}}c@{\hspace{0.5cm}}c}
 Atom & $x/a$ & $y/b$ & $z/c$ & $U_{\rm iso}$ (\r A$^2$) \\
Fe & 0.5011(2)  & 0.5040(3)   &  0.1149(3)   & 0.0086(6) \\
 Sn1 & 0 & 0 & 0.0884(3)  &  0.0089(5) \\
 Sn2 & 0  & 0 &  0.3338(3) & 0.0098(5) \smallskip\\\hline
\end{tabular}
\label{Fe32_26GPa}
\end{table}

\section*{Supplementary references}
\begin{small}

\begin{enumerate}[label={$\left[S\arabic*\right]$}]
\hypertarget{cite}\item Tuniz, M. $et$ $al.$ ``Dynamics and resilience of the unconventional charge density wave in ScV$_6$Sn$_6$ bilayer kagome metal", \href{https://doi.org/10.1038/s43246-023-00430-y} {Commun. Mater.  \textbf{4}, 1 (2023)}.
\item Wang, Z. X. $et$ $al.$ ``Unconventional charge density wave and photoinduced lattice symmetry
change in the kagome metal CsV$_3$Sb$_5$ probed by time-resolved spectroscopy", \href{https://doi.org/10.1103/PhysRevB.104.165110} {Phys. Rev. B  \textbf{104}, 165110 (2021)}.
\item Liu, Y.$et$ $al.$ ``Visualizing electron–phonon and anharmonic phonon–phonon coupling in
the kagome ferrimagnet GdMn$_6$Sn$_6$", \href{https://doi.org/10.1063/5.0152116} {Applied Physics Letters  \textbf{122}, 251901 (2023)}.
\item Birch, F. ``Finite elastic strain of cubic crystals", \href{https://link.aps.org/doi/10.1103/PhysRev.71.809} {Phys. Rev. \textbf{71}, 809 (1947)}.
\item Holzapfel, W. ``Equations of state for solids under strong compression", \href{https://doi.org/10.1524/zkri.216.9.473.20346} {Zeitschrift f\"{u}r Kristallographie - Crystalline Materials  \textbf{216}, 473 (2001)}.
\end{enumerate}

\end{small}

\end{document}